\documentstyle[12pt,aaspp4,epsf]{article}
\newcommand\ltorder{\mathrel{\raise.3ex\hbox{$<$}\mkern-14mu
             \lower0.6ex\hbox{$\sim$}}}
\newcommand\gtorder{\mathrel{\raise.3ex\hbox{$>$}\mkern-14mu
             \lower0.6ex\hbox{$\sim$}}}

\begin{document}

\title{Structure of Dark Matter Halos From Hierarchical Clustering:\\
    II. Universality and Self-Similarity in Cluster-Sized Halos} 

\author{Toshiyuki Fukushige}

\affil{Department of General Systems Studies,\\
College of Arts and Sciences, University of Tokyo,\\
 3-8-1 Komaba, Meguro-ku, Tokyo 153-8902, Japan}

\author{Junichiro Makino}

\affil{Department of Astronomy,\\
School of Sciences, University of Tokyo,\\
 7-3-1 Hongo, Bunkyo-ku, Tokyo 117-0033, Japan}


\begin{abstract}

We investigate the structure of the dark matter halo formed in three
different cold dark matter scenarios.  We performed N-body simulations
of formation of 13 cluster-sized halos.  In all runs, density cusps
proportional to $r^{-1.5}$ developed at the center.  This result was
independent of the cosmological models we simulated.  We could not
reproduce the cusp shallower than $r^{-1.5}$, which was obtained in some
of previous studies.  We also found that in all runs the density
structure evolves in a self-similar way, even in $\Omega\ne 1$ universes.  
These results show that the formation of structural form is a
process decoupled from a background cosmology. 

\end{abstract}


\section{Introduction}

The structure of dark matter halos formed through disspationless
hierarchical clustering from cosmological initial setting has been
explored by many researchers since the "finding" of the universal
profile by Navarro, Frenk and White (1996, 1997, hereafter NFW).  NFW
performed $N$-body simulations of the halo formation and found that the
density profiles of dark matter halos were expressed well by a simple
formula,
\begin{equation}
\rho={\rho_0 \over (r/r_0)(1+r/r_0)^2}
\end{equation}
where $\rho_0$ is a characteristic density and $r_0$ is a scale radius. 
They argued that the profile of dark matter halos have the same shape,
independent of the halo mass, the initial density fluctuation spectrum
or the value of the cosmological parameters. 

Although the NFW results have been confirmed regarding the logarithmic
slope shallower than isothermal near the center and $-3$ in
the outskirt, disagreements concerning the slope of the central region
were reported by subsequent studies in which higher-resolution
simulations were performed.  The disagreements are summarized into the
following two kinds. 

One is that the slope at the center is steeper than that in the NFW
result.  Fukushige and Makino (1997, hereafter FM97) performed
a simulation with 768k particles, while previous studies employed $\sim
20$k, and found that the galaxy-sized halo in their simulation has a
cusp steeper than $\rho\propto r^{-1}$.  Moore et al.  (1998, 1999
hereafter M99) and Ghigna et al.  (2000) performed simulations with up
to 4M particles and obtained the results that the profile has a cusp
proportional to $r^{-1.5}$ both in galaxy-sized and cluster-sized halos. 
They proposed the modified universal profile (hereafter, the M99
profile),
\begin{equation} 
\rho = {\rho_0 \over (r/r_{\rm s})^{1.5}[1+(r/r_{\rm s})^{1.5}]}.  
\label{eqmoo} 
\end{equation}
Klypin et al.  (2001) also obtained the results the the slope at the
center can be approximated by $r^{-1.5}$, though they argued that the
NFW fit is still good up to their resolution limit. 

The second is that the slope at the center is not universal.  Jing and
Suto (2000) performed a series of $N$-body simulations and concluded
that the power of the cusp depends on mass, in contradiction to the
claims by earlier studies.  It varies from $-1.5$ for galaxy mass halo
to $-1.1$ for cluster mass halo.  Also, many analytical studies argued
that the halo profile should depend on the power spectrum of initial
density fluctuation.  For example, Hoffman and Shaham (1985) and Syer
and White (1998) predicted the slope, $3(n+3)/(n+4)$ and $3(n+3)/(n+5)$,
respectively, where $n$ is the effective power-law index of the power
spectrum. 
  
In the first paper of this series (Fukushige and Makino 2001, hereafter
Paper I), we performed simulations of 12 halos with the mass ranging
from $6.6\times 10^{11}M_{\odot}$ to $8.0\times 10^{14}M_{\odot}$.  We
found that, in all runs, the halos have density cusps proportional to
$r^{-1.5}$ developed at the center.  This result means that the density
structure is universal in the sense that it is independent of the halo
mass.  We also found that the density structure evolves in a
self-similar way as the central cusp grows outward with keeping the
density near the center unchanged. 

This paper is a follow up of Paper I.  In this paper, we focus on the
dependence of the halo profile on the cosmological model.  We simulate
halos in three conventional cold dark matter models: Standard, Lambda
and Open CDM models, while simulations in Paper I was performed only in
the standard CDM model.  We performed $N$-body simulations of formation
of 13 cluster-sized dark matter halos using Barnes-Hut treecode and a
special-purpose computer GRAPE-5 (Kawai et al.  2000).  This is the
first study that investigates the dependence on the different CDM models
with the mass resolution and number of samples enough to discuss the
slope of the cusp, though several simulations were performed in the same
context (NFW, Thomas et al.  1998, Huss, Jain, and Steinmetz 1999). 

The structure of this paper is as follows.  In section 2, we describe
the model of our $N$-body simulation.  In section 3, we present the
results of simulation.  Section 4 is for conclusion and discussion. 

\section{Simulation Method}

We simulated the formation of the dark matter halos using
"re-simulation" method, which has been the standard method for the
simulation of halo formation since NFW.  In this method, we first
perform large scale cosmological simulations in order to identify the
halo candidates.  We trace back to the initial condition of the large
scale simulation, and express the halo candidate with larger number of
particles by adding a shorter wavelength perturbation.  Then, we
resimulate the halo candidates. 

We used three cosmological models listed in Table \ref{tab1}: Standard,
Open and Lambda Cold Dark Matter models (SCDM, OCDM, and LCDM).  Here,
$\Omega_0$ is the density parameter, $\lambda_0$ is the dimensionless
cosmological constant, $H_0 = 100 h \, {\rm km}\cdot{\rm s}\cdot{\rm
Mpc}^{-1}$ at the present epoch, and $z_i$ is the initial redshift.  
The amplitudes of the power spectrum in CDM models were
normalized using the top-hat filtered mass variance at $8 h^{-1}$ Mpc
according to the cluster abundance (Kitayama \& Suto 1997). 

The large scale cosmological simulations were performed with $1.1\times
10^6$ particles in a sphere of 100Mpc radius.  The procedure for setting
the initial condition were the same as those used in Fukushige and Suto
(2001).  We regard a spherical overdensity region around a local
potential minimum within $r_{\rm v}$ as a candidate halo.  We define the
radius $r_{\rm v}$ such that the spherical overdensity inside is
178$\Omega_0^{0.3}$ times the critical density for SCDM and OCDM model,
and 178$\Omega_0^{0.4}$ times for LCDM model (Eke, Cole, Frenk 1996). 

We selected 13 halos from the candidate halos catalog, which are
summarized in Table \ref{tab2}.  We first selected the most massive halo
in each CDM model, and then selected the halos at random from halo
candidates containing more than 1000 particles.  We expressed a region
within $5r_{\rm v}$ from the center of the halo at $z=0$ in the
cosmological simulation with larger number of particles.  We placed
particles whose mass is the same as that in the cosmological simulation
in a sphere of $\sim$ 50Mpc radius surrounding the high resolution
region in order to express the external tidal field.  The total number
of particles, $N$, is listed in Table \ref{tab2}.  As a result, the mass
ratio between the high resolution particles and surrounding particles
becomes rather large, $256 \sim 2048$.  In order to prevent the
contamination of heavy particles into the halo, we set the boundary to
be a rather large value, $5r_{\rm v}$.  No heavy particle entered
within $r_{\rm v}$ throughout all simulations. 

We integrated the system directly in the physical coordinates for both
the cosmological and halo simulations (as in FM97, Paper I).  We used a
leap-flog integrator with shared and constant timestep.  The step size
for the cosmological simulation is $\Delta t/t_{\rm H}=1/1024$ and that
for the halo simulation is listed in Table \ref{tab2}.  The code for the
time integration is the same as that in Fukushige and Suto (2001).  We
used the usual Plummer softening.  The gravitational softening
$\varepsilon$ is constant in the physical coordinates and the length is
5kpc for the cosmological simulation and 1kpc for all halo simulations. 

The force calculation is done with the Barnes-Hut tree code (Barnes \&
Hut 1986, Makino 1991) on GRAPE-5 (Kawai et al.  2000), a
special-purpose computer designed to accelerate $N$-body simulations. 
For most simulations, we used the GRAPE-5 system at the Astronomical
Data Analysis Center of the National Astronomical Observatory, Japan. 
We used the opening parameter $\theta = 0.4$ for the cosmological
simulation and $\theta=0.5$ for the halo simulation.  The simulations
presented below required, for example in Run L6, $\sim 80$ secs per
timestep, and thus one run (16k timesteps) was completed in 370 CPU
hours with a GRAPE-5 board connected to a host workstation with Alpha
21264 CPU (833MHz). 

\begin{table}[h]
\caption{Cosmological Models \label{tab1}}
\begin{tabular}{lcccccc}
\hline
\hline
Model & $\Omega_0$ & $\lambda_0$ & $h$ & $\sigma_8$ 
& $\rho_{\rm crit}$($M_{\odot}/pc^{3}$) & $z_i$  \\
\hline
SCDM & 1.0 & 0.0 & 0.5 & 0.6 & $6.9 \times 10^{-8}$ & 24.0 \\
LCDM & 0.3 & 0.7 & 0.7 & 1.0 & $1.4 \times 10^{-7}$ & 32.3 \\
OCDM & 0.3 & 0.0 & 0.7 & 1.0 & $1.4 \times 10^{-7}$ & 32.3 \\
\hline
\end{tabular}
\end{table}

\begin{table}[h]
\caption{Simulation Models \label{tab2}}
\begin{tabular}{lc|ccr|ccc}
\hline
\hline
Model & Run & $M_{\rm v}$($M_{\odot}$) & $r_{\rm v}$ (Mpc) 
& $N_{\rm v}$ & $m$ ($10^{8}M_{\odot}$)& $\Delta t$($10^{6}$yr) & $N$ \\
\hline
SCDM & S1 & $8.67\times 10^{14}$ & 2.62 & 1676525 & 5.17 & 1.57 & 5074432 \\
     & S2 & $5.46\times 10^{14}$ & 2.21 & 1056312 & 5.17 & 1.57 & 3523844 \\
     & S3 & $3.68\times 10^{14}$ & 1.97 & 1421930 & 2.58 & 1.57 & 3478480 \\
     & S4 & $3.58\times 10^{14}$ & 1.91 & 1383674 & 2.58 & 1.57 & 4104120 \\
\hline
LCDM & L1 & $7.83\times 10^{14}$ & 2.34 & 1288779 & 6.08 & 0.82 & 3624848 \\
     & L2 & $5.32\times 10^{14}$ & 2.08 &  875058 & 6.08 & 1.63 & 4360512 \\
     & L3 & $3.97\times 10^{14}$ & 1.85 & 1306187 & 3.04 & 0.82 & 3066944 \\
     & L4 & $2.17\times 10^{14}$ & 1.52 & 1425526 & 1.52 & 0.82 & 3536640 \\
     & L5 & $2.15\times 10^{14}$ & 1.52 &  707569 & 3.04 & 1.63 & 2058140 \\
     & L6 & $1.83\times 10^{14}$ & 1.43 & 1809105 & 1.01 & 0.82 & 5458688 \\
\hline
OCDM & O1 & $8.58\times 10^{14}$ & 2.34 & 1411523 & 6.08 & 0.68 & 3711232 \\
     & O2 & $4.27\times 10^{14}$ & 1.86 &  702022 & 6.08 & 1.37 & 1748480 \\
     & O3 & $2.18\times 10^{14}$ & 1.47 &  717056 & 3.04 & 1.37 & 2148736 \\
\hline
\end{tabular}
\end{table}

\section{Results}

\subsection{Snapshots}

\begin{figure}
\begin{center}
{\leavevmode
\epsfxsize=15cm
\epsfbox{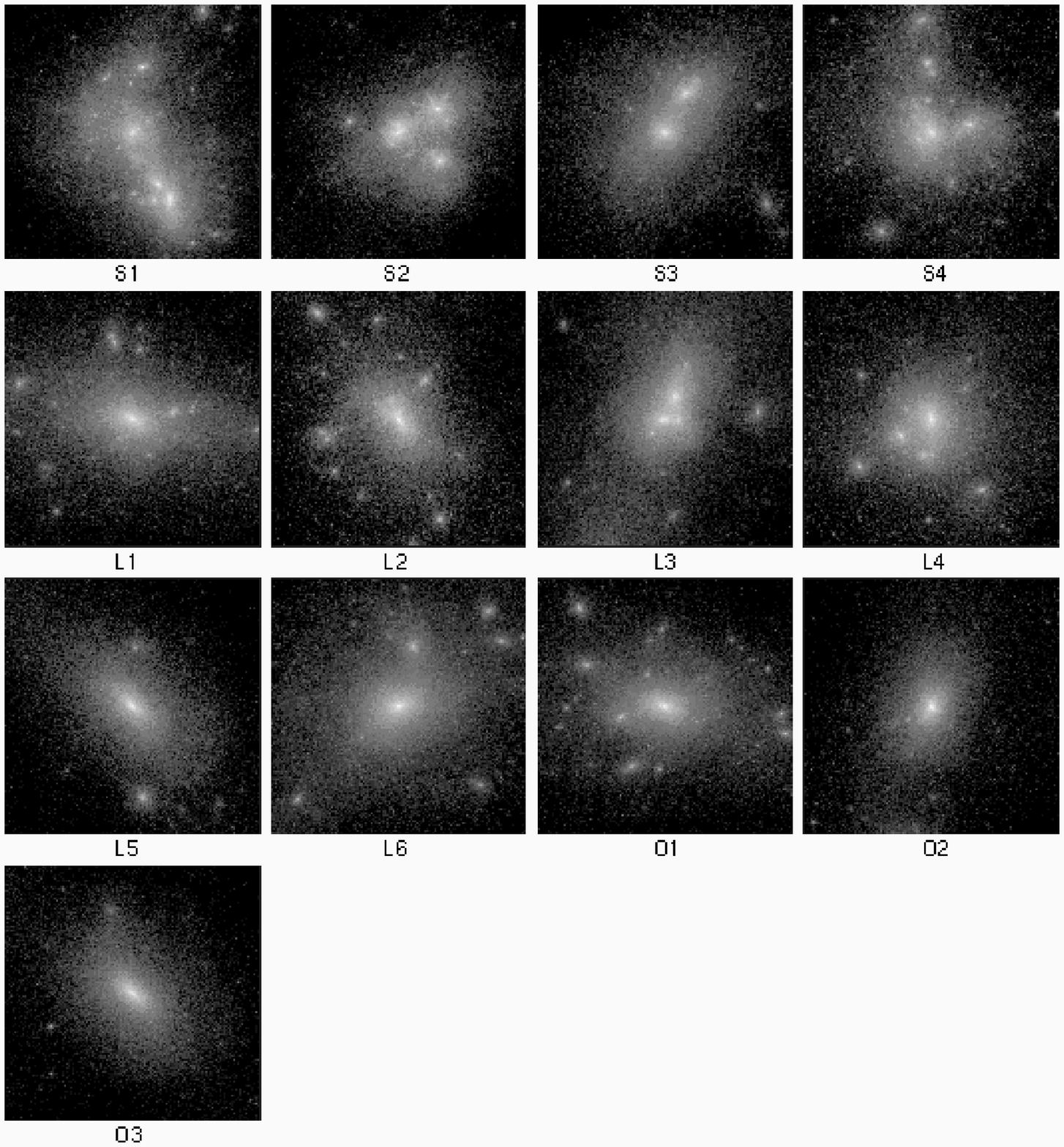}
\caption{Snapshots from all Runs at $z=0$.  The length of the side for
each panel is equal to $2r_{\rm v}$.\label{fig1}}
}
\end{center}
\end{figure}

Figure \ref{fig1} shows the particle distribution for all runs at $z=0$. 
The length of the side for each panel is equal to $2r_{\rm v}$.  For
these plots, we shifted the origin of coordinates to the position of the
potential minimum.  In Table \ref{tab2}, we summarized the radius
$r_{\rm v}$, the mass $M_{\rm v}$, and the number of particles $N_{\rm
v}$ within $r_{\rm v}$ at $z=0$. 

\subsection{Density Profile}

Figure \ref{fig2} shows the density profiles for all runs at $z=0$.  For
Run O2, we plot the density profile at $z=0.026$ because the merging
process occurs just near the center of halo at $z=0$.  The position of
the center of the halo was determined using the potential minimum and
the density is averaged over each spherical shell whose width is
$\log_{10}(\Delta r)=0.0125$.  For the illustrative purpose, the
densities are shifted vertically. 

In this figure, we plot the densities in the thick lines if the two
criteria introduced in Paper I; (1) $t_{\rm rel}(r)/t > 3$ and (2)
$t_{\rm dy}(r)/\Delta t > 40$, are satisfied, where $t_{\rm rel}(r)$ is
the local two-body relaxation time and $t_{\rm dy}(r)$ is the local
dynamical time.  This means that the densities in the central region
plotted in the thin lines are influenced by the numerical artifacts
(mainly two-body relaxation), and they are not reliable.  In the
following discussion, we only use the densities plotted in the thick
lines. 
  
In all runs we can see the central density cusps approximately
proportional to $r^{-1.5}$.  In other words, the power of the cusp is
$-1.5$ and is independent of cosmological models we simulated.  The
shallowing of the power-law index of the inner cusp observed in the LCDM
runs by Jing and Suto (2000) was not reproduced in our LCDM runs. 

Moreover, the density profiles are in good agreement to the profile
given by equation (\ref{eqmoo}) (the M99 profile) in all runs.  We set
the scale radii $r_0$ as 0.8, 0.5, 0.4, 0.5 Mpc for S1..S4 runs, 0.6,
0.5, 0.5, 0.35, 0.3, 0.3 Mpc for L1..L6 runs, and 0.35, 0.35, 0.25 Mpc
for O1..O3 runs.  The agreement is very good for the inner region.  In
the outer region the agreement is not very good simply because the outer
profile shows large fluctuations caused by individual infalling halos. 
The degree of the agreement is better in LCDM and OCDM model than in
SCDM model.  This is because the halo in LCDM and OCDM model is
typically formed earlier and it is dynamically quiet around $z\sim 0$. 

Figure \ref{fig3} shows the scale densities of the profile $\rho_0$
[equation (\ref{eqmoo})] and the concentration parameter $c \equiv
r_{\rm v}/r_0$ as a function of total mass.  These values in S/LCDM
models are consistent to those obtained by M99 and Jing and Suto (2000). 
We can see a tendency that that cluster-size halos in OCDM models are
more compact than that of their S/LCDM counterparts of the same mass. 

\subsection{Self-Similarity}
\label{secself}

In the SCDM simulations of Paper I, we found that the density profile
evolves in a self-similar way as the central cusp grows outward keeping
the density near the center unchanged (illustrated in Figure 14 of Paper
I).  However, it is not clear whether the density profile evolves in the
same way in the LCDM and OCDM model, because neither of models
retains the scale-free nature of the SCDM model. 

If the evolution is self-similar, we can write the density as 
\begin{eqnarray}
\rho(r,t) &=& \rho_{\dagger}(M)\rho_{\ast}(r_{\ast})\\
r_{\ast} &=& r/r_{\dagger}(M)
\end{eqnarray}
In Paper I, we obtained the self-similar variables, $\rho_{\dagger}$ and 
$r_{\dagger}$, as 
\begin{eqnarray} 
\displaystyle 
\rho_{\dagger}(M) &=& \left({M\over 10^{14}M_{\odot}}\right)^{n\over 3+n}\label{eq1}\\
r_{\dagger}(M) &=& \left({M\over 10^{14}M_{\odot}}\right)^{1\over 3+n}\label{eq2}
\end{eqnarray} 
assuming that the halo having a $r^{n}$ cusp grows outward in a
self-similar way keeping the density of central cusp region constant and
the fraction of mass in the cusp to total mass, $M$, is constant.  Here
$n$ is the power of the cusp. 

In Figure \ref{fig4} we plot the scaled density $\rho_{\ast}$ as a
function of $r_{\ast}$.  We plot three profiles at different value of
the redshifts $z$ for six runs.  We set $n$ to be $-1.5$ and use $M_{\rm
v}$ as the total mass.  In this figure, we can see that the density
profiles of the same halo at different times show good agreement to each
other, which means that the density structure evolves self-similarly. 
Therefore, Figure \ref{fig4} demonstrates that our assumption of the
self-similarity is justified not only in the SCDM model, but also in
LCDM and OCDM models. 

Finally, we consider whether halos of different masses are on the
self-similar evolution track or not.  If a halo evolves self-similarly,
scaled scale densities $\rho_{\ast 0}\equiv \rho_{0}/\rho_{\dagger}$
should be constant during the evolution.  Using equation (\ref{eq1}), the
scale density $\rho_{0}$ is proportional to $M^{-1}$.  In Figure
\ref{fig3}, the dashed line indicates $\rho_0 \propto M^{-1}$.  If the
halo evolve self-similarly, the evolution track is along the dashed line
in the $\rho_0-M$ plot.  We can see a tendency that at least for O/LCDM
models, halos do not distribute along the self-similar evolution track,
in other words, small halos cannot grow to large halos even if we wait
longer. 

\section{Conclusion and Discussion}

We performed $N$-body simulations of dark matter halo formation in three
CDM models: Standard, Lambda, and Open CDM models.  We simulated 13 halos
whose mass range is $1.8\times 10^{14}M_{\odot}$ to $8.6\times
10^{14}M_{\odot}$.  We used a widely adapted "re-simulation" method to
set up initial conditions of halos and include the external tidal
field. 

Our main conclusions are: 

\begin{itemize}

\item[(1)] 
We found that, in all runs, the final halos have density cusps
proportional to $r^{-1.5}$, and the profiles show good agreement with
the M99 profile, regardless of the cosmological models. 

\item[(2)] 
In all runs, the density profile evolves self-similarly.  This is also
independent of the cosmological models we simulated. 

\end{itemize}

\begin{figure}
\begin{center}
{\leavevmode
\epsfxsize=17cm
\epsfbox{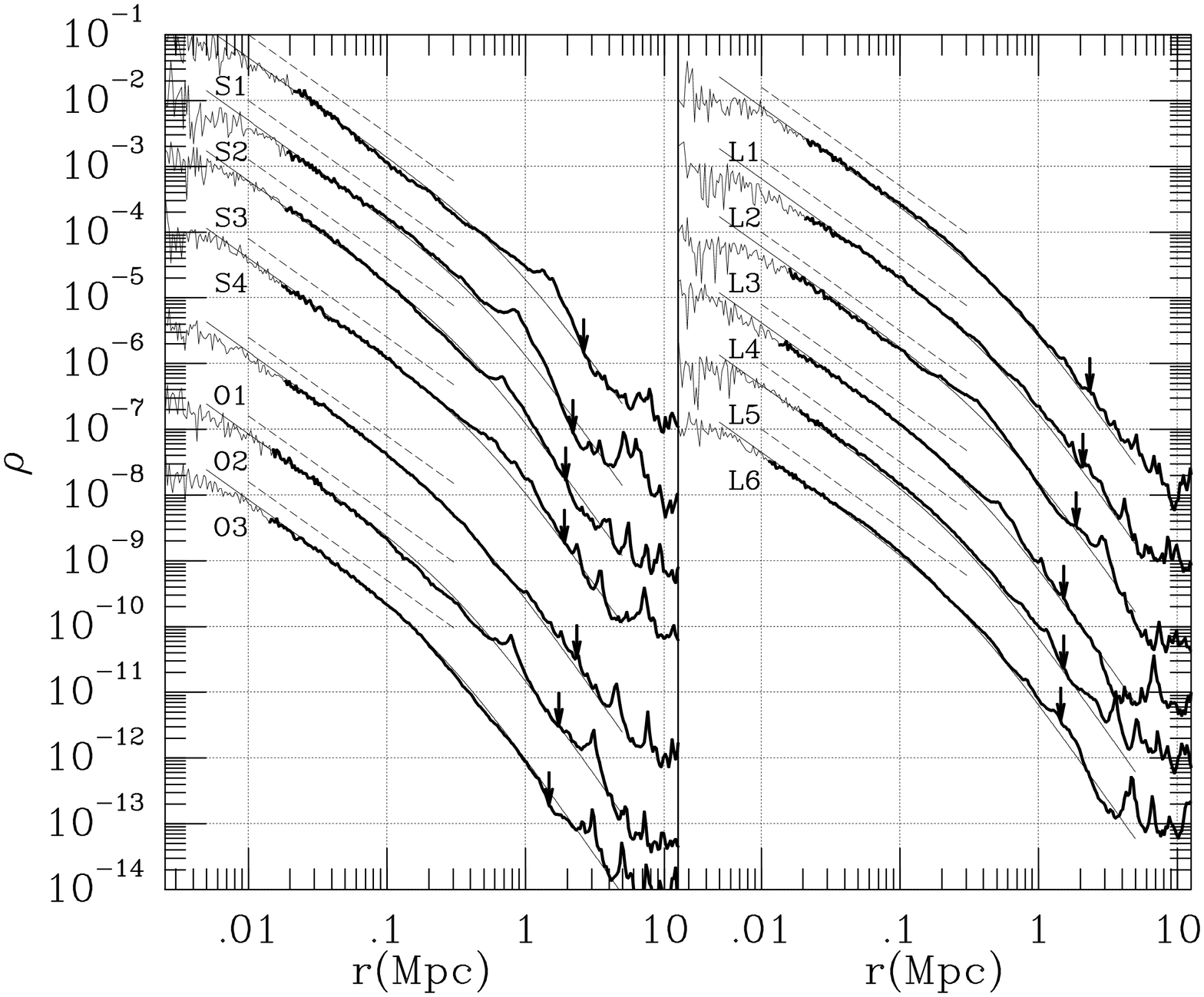}
\caption{
Density profiles of the halos for all Runs at $z\simeq 0$.  Only the
densities plotted in the thick lines satisfy the accuracy criteria in
section 3.2. Those plotted in the thin lines are influenced
by numerical artifacts.  The unit of density
is $M_{\odot}$/pc$^3$.  The labels on the left of the profiles indicate
the run name.  The profiles except for Runs S1 are vertically shifted
downward by 1, 2, 3, 5, 6, 7 dex for Runs S2, S3, S4, O1, O2, O3, and by
1...6 for Runs L1...L6, respectively.  The arrows indicate $r_{\rm v}$. 
The thin dashed lines indicate the densities proportional to $r^{-1.5}$. 
The thin solid curves indicate the density profile given by equation
(\ref{eqmoo})(the M99 profile). 
\label{fig2}}
}
\end{center}
\end{figure}

\begin{figure}
\begin{center}
{\leavevmode
\epsfxsize=15cm
\epsfbox{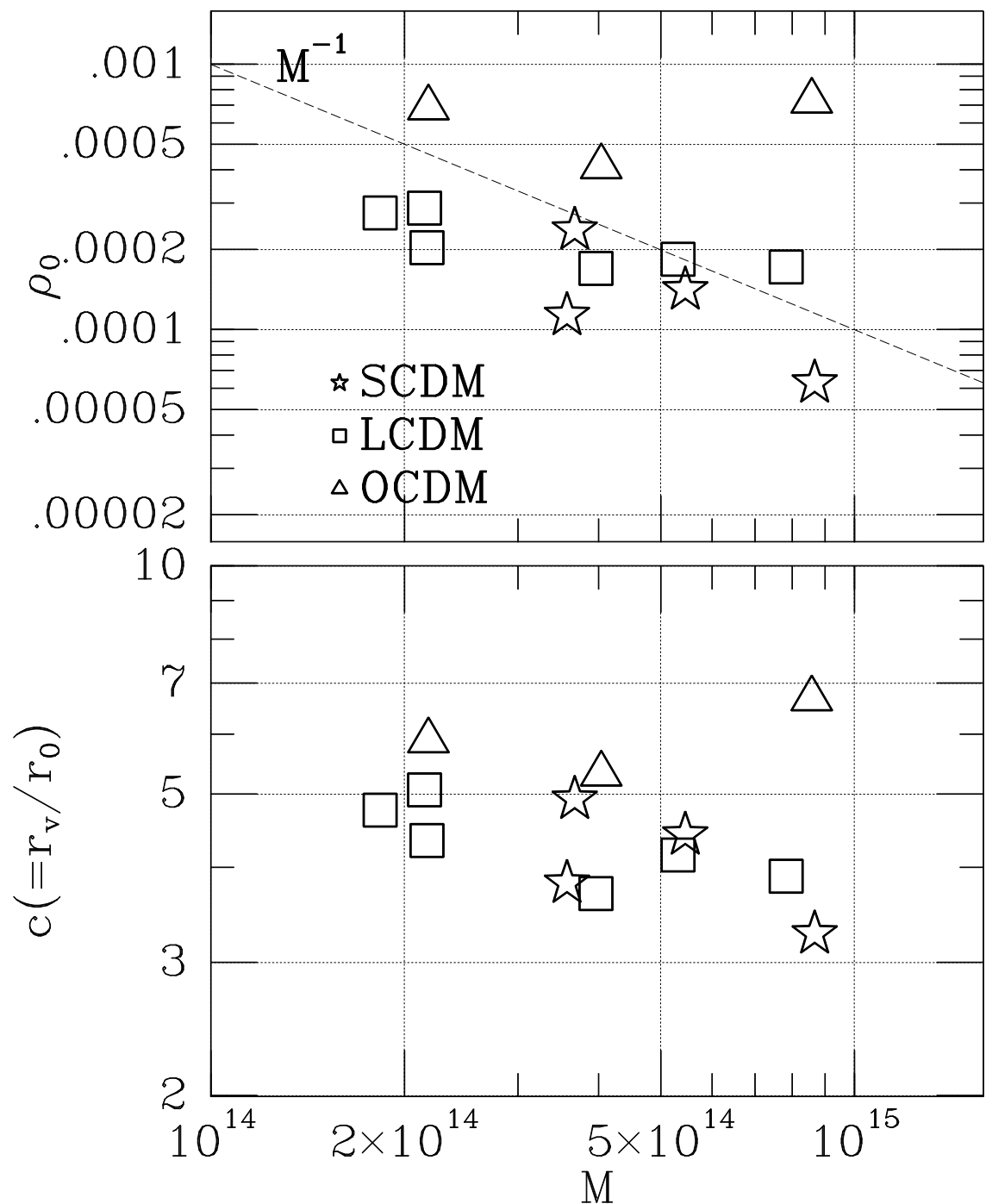}
\caption{Scale density $\rho_{0}$ in the unit of $M_{\odot}$/pc$^3$ and
concentration parameter $c=r_{\rm v}/r_{0}$ as a function of total mass
in solar mass.  The star, square, and triangle symbols show those for
SCDM, LCDM, and OCDM models, respectively. The dashed line 
indicates the self-similar evolutionary track discussed in section \ref{secself}.
\label{fig3}}
}
\end{center}
\end{figure}

\begin{figure}
\begin{center}
{\leavevmode
\epsfxsize=13cm
\epsfbox{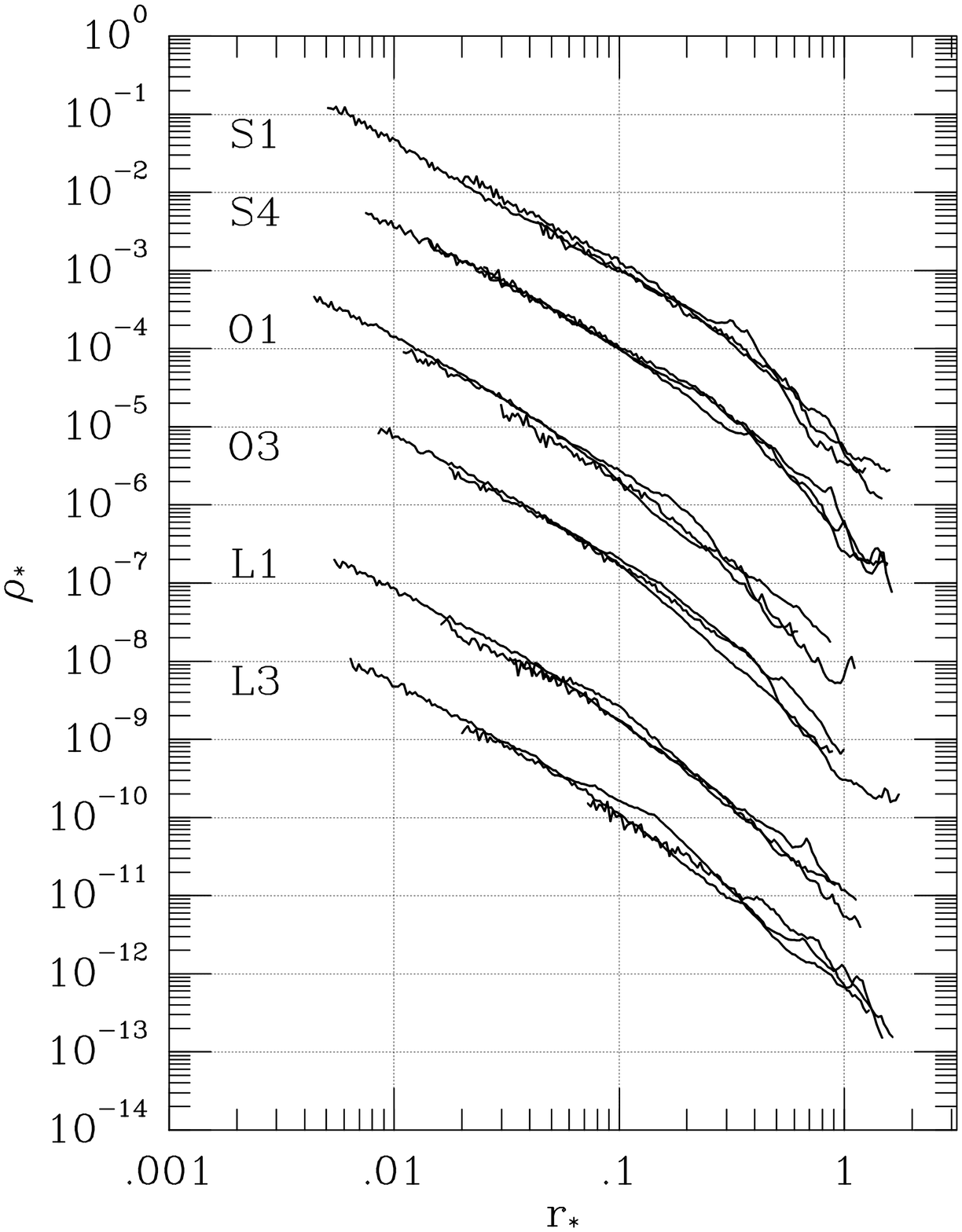}
\caption{
Self-similar evolution of the density profile.  The scaled densities
$\rho_{\ast}$ are plotted as a function of the scaled radius $r_{\ast}$. 
The profile for Run S4, O1, O3, L1 and L3 are vertically shifted
downward by 1, 3, 4, 6, and 7 dex, respectively.  The densities within
2$r_{\rm v}$ at each redshift are plotted. 
\label{fig4}}
}
\end{center}
\end{figure}

There are some claims that the innermost slope should converge not to
$r^{-1.5}$ but to a shallower one (e.g.~Taylor, Navarro 2001).  Indeed,
if we do not pay attention to numerical artifacts, we may see in our
simulation results (Figure \ref{fig2}) that the innermost slopes of all
runs become shallower than $-1.5$.  However, Figure \ref{fig2} shows
that the inner region where the slope are shallower than $r^{-1.5}$ is
not reliable because in this region the accuracy criteria are not
satisfied.  The numerical artifact which makes the cusp shallower is
mainly the two-body relaxation effect in this region (see Paper I). 
Therefore, we emphasize that discussions based on simulation results
without careful analysis of the influence by the two-body relaxation
effects are misleading. 

In our LCDM run, we could not reproduce the shallowing of the power-law
index of the inner cusp observed in the LCDM runs by Jing and Suto
(2000).  This difference could be due to the smoothing by two-body
relaxation in their cluster-sized halos.  In this paper, we show that
the density profile within $\sim 0.01r_{200}$ smoothed by the two-body
relaxation.  The density at $0.01r_{200}$ and the mass resolution in
their cluster-sized halo are similar to those in ours.  The density
profile in their simulations within $0.01r_{200}$, at which the profile
begins to depart from $r^{-1.5}$ inward, could be affected by the
two-body relaxation.  If their simulations had been performed with
higher mass resolution, the slope might have approach to $-1.5$.  The
tendency can be already seen in their galaxy-sized halo.  All halos have
the cusp proportional to $r^{-1.5}$ in their galaxy-sized halos whose
central densities at $0.01r_{200}$ are almost similar to those of their
cluster-sized run and mass resolutions are about 100 times higher. 

Our results show that the NFW's claim concerning to the universality is
certainly valid.  On the other hand, the density profile obtained are not
in agreement with the NFW profile at the central region.  Although it is
important to find the convergence slope by further simulations, a simple
explanation would be required for our final understanding of the
universal profile.  At present, we don't fully understand why the
profile is universal and/or why the power of central cusp is $-1.5$,
which we will address it in the future study. 

\acknowledgements

We are grateful to Atsushi Kawai and Eiichiro Kokubo for their helps in
preparing the hardware and software environment of the GRAPE-5 system,
and to Yasushi Suto, Atsushi Taruya, and Masamune Oguri for many helpful
discussions.  We gratefully acknowledge the use of the initial condition
generator in the publicly available code {\it Hydra} (1995) developed by
H.M.P.Couchman, P.A.Thomas, and F.R.Pearce.  Most of numerical
computations were carried out on the GRAPE system at ADAC (the
Astronomical Data Analysis Center) of the National Astronomical
Observatory, Japan.  This research was partially supported by the
Research for the Future Program of Japan Society for the Promotion of
Science, grant no.  JSPS-RFTP 97P01102.

\begin{references}

\reference{}
Barnes, J. E., \& Hut, P. 1986, Nature, 824, 446

\reference{}
Couchman, H. M. P., Thomas, P. A., \& Pearce, F. R. 1995, ApJ, 452, 797

\reference{}
Eke, V. R., Cole, S., \& Frenk C. S. 1996, MNRAS, 282, 263

\reference{}
Fukushige, T., \& Makino, J. 1997, ApJ, 477, L9

\reference{}
Fukushige, T., \& Makino, J. 2001, ApJ, in press (Paper I)

\reference{}
Fukushige, T., \& Suto, Y. 2001, ApJL, in press

\reference{}
Ghigna, S., Moore, B., Governato, F., Lake, G., Quinn, T., \&
Stadel, J. 2000, ApJ, 544, 616

\reference{}
Hoffman, Y., \& Shaham, J. 1985, ApJ, 297, 16

\reference{}
Huss, A., Jain, B., \& Steinmetz, M. 1999, MNRAS, 308, 1011

\reference{}
Jing, Y. P., \& Suto, Y. 2000, ApJ, 529, L69

\reference{}
Kawai, A., Fukushige, T., Makino, J., \& Taiji, M. 2000, PASJ, 52, 659

\reference{}
Kitayama, T., \& Suto, Y. 1997, ApJ, 490, 557

\reference{}
Klypin, A., Kravtson, A. V,, Bullock, J. S., \& Primack, J. R. 2001, 
ApJ, 554, 903

\reference{}
Makino, J. 1991, PASJ, 43, 621

\reference{}
Moore, B., Governato, F., Quinn T., Statal, J., \& Lake, G. 1998, ApJ, 499, L5

\reference{}
Moore, B., Quinn T., Governato, F., Statal, J., \& Lake, G. 1999, MNRAS, 310, 1147

\reference{}
Navarro, J. F., Frenk, C. S., \& White, S. D. M., 1996, ApJ, 462, 563 

\reference{}
Navarro, J. F., Frenk, C. S., \& White, S. D. M., 1997, ApJ, 490, 493

\reference{}
Syer, D., \& White, S. D. M. 1998, MNRAS, 293, 337

\reference{}
Thomas, P., et al. 1998, MNRAS, 296, 1061

\end{references}
\end{document}